Nuclear structure of $^{140}$Te with N = 88: Structural symmetry and asymmetry in Te isotopes with respect to the double-shell closure Z = 50 and N = 82

C.-B. Moon,[1,*] P. Lee,[2] C. S. Lee,[2] A. Odahara,[3] R. Lozeva,[4] A. Yagi,[3] F. Browne,[5,6] S. Nishimura,[5] P. Doornenbal,[5] G. Lorusso,[5] P.-A. Söderström,[5] T. Sumikama,[7] H. Watanabe,[5] T. Isobe,[5] H. Baba,[5] H. Sakurai,[5] R. Daido,[3] Y. Fang,[3] H. Nishibata,[3] Z. Patel,[5,8] S. Rice,[5,8] L. Sinclair,[5,9] J. Wu,[5,10] Z. Y. Xu,[11] R. Yokoyama,[12] T. Kubo,[5] N. Inabe,[5] H. Suzuki,[5] N. Fukuda,[5] D. Kameda,[5] H. Takeda,[5] D. S. Ahn,[5] D. Murai,[13] F. L. Bello Garrote,[14] J. M. Daugas,[15] F. Didierjean,[4] E. Ideguchi,[16] T. Ishigaki,[3] H. S. Jung,[17], T. Komatsubara,[5] Y. K. Kwon,[18] S. Morimoto,[3] M. Niikura,[5,11] I. Nishizuka,[6] and K. Tshoo[18]

[1] *Department of Display Engineering, Hoseo University, Chung-Nam 336-795, Korea*
[2] *Department of Physics, Chung-Ang University, Seoul, 156-756, Korea*
[3] *Department of Physics, Osaka University, Osaka 560-0043, Japan*
[4] *IPHC, CNRS, IN2P3 and University of Strasbourg, F-67037 Strasbourg Cedex 2, France*
[5] *RIKEN Nishina Center, Wako, Saitama 351-0198, Japan*
[6] *School of computing engineering and mathematics, University of Brighton BN2 4JG, United Kingdom*
[7] *Department of Physics, Tohoku University, Sendai, Miyagi 980-8578, Japan*
[8] *Department of Physics, University of Surrey, Guildford, GU2 7XH, United Kingdom*
[9] *Department of Physics, University of York, Heslington, York YO10 5DD, United Kingdom*
[10] *School of Physics and State key Laboratory of Nuclear Physics and Technology, Peking University, Beijing 100871, China*
[11] *Department of Physics, University of Tokyo, Tokyo 113-0033, Japan*
[12] *Center for Nuclear Study, University of Tokyo, RIKEN Campus, Wako, Saitama 351-0198, Japan*
[13] *Department of Physics, Rikkyo University, Tokyo 172-8501, Japan*
[14] *Department of Physics, University of Oslo N-0316, Norway*
[15] *CEA, DAM, DIF, F-91297 Arpajon Cedex, France*
[16] *RCNP, Osaka University, Osaka 567-0047, Japan*
[17] *Wako Nuclear Science Center (WNSC), Institute of Particle and Nuclear Studies (IPNS), High Energy Accelerator Research Organization(KEK), Wako, Saitama 351-0198, Japan*
[18] *Rare Isotope Science Project, Institute for Basic Science, Daejeon 305-811, Korea*

We study for the first time the internal structure of $^{140}$Te through the beta($\beta$)-delayed gamma($\gamma$)-ray spectroscopy of $^{140}$Sb. The very neutron-rich $^{140}$Sb, Z = 51 and N = 89, ions were produced by the in-flight fission of $^{238}$U beam on a $^{9}$Be target at 345 MeV per nucleon at the Radioactive Ion Beam Factory, RIKEN. The half-life and spin-parity of $^{140}$Sb are reported as 124(30) ms and (4$^-$), respectively. In addition to the excited states of $^{140}$Te produced by the $\beta$-decay branch, the $\beta$-delayed one-neutron and two-neutron emission branches were also established. By identifying the first 2$^+$ and 4$^+$ excited states of $^{140}$Te, we found that Te isotopes persist their vibrator character with E(4$^+$)/E(2$^+$) = 2. We discuss the distinctive features manifest in this region, such as valence neutron symmetry and asymmetry, revealed in pairs of isotopes with the same neutron holes and particles with respect to N = 82.

*cbmoon@hoseo.edu

The shell structure of the atomic nucleus is one of the cornerstones for a comprehensive understanding of the many-body quantum mesoscopic system. Fundamental characteristics of nuclear structure are best represented by systematic changes of experimental observables across the nuclear chart [1-4]. Especially illuminating are the systematics of the first $2^+$ excited states of isotopic and isotonic chains which span the major shell closures. Figure 1 depicts such systematics of the even-even $46 \leq Z \leq 54$ isotopes that show clear correlations between the $_{52}$Te - $_{48}$Cd and $_{54}$Xe - $_{46}$Pd isotopic chains [4].

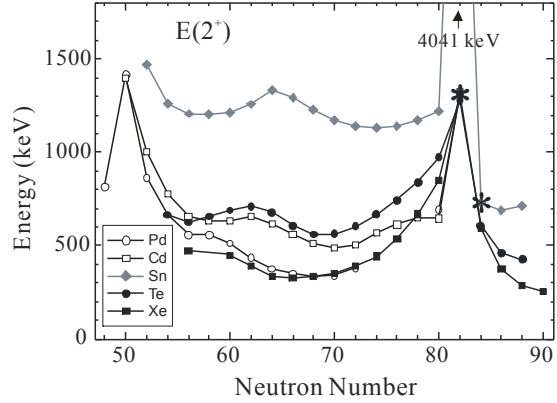

Fig. 1. Systematics for the first $2^+$ excited states in the nuclides around Sn (Z = 50) as a function of neutron number [4]. Data are primarily from [5], [6] for $^{126,128}$Pd, [7] for $^{136,138}$Sn, [8] for $^{138}$Te, and the present work for $^{140}$Te. For discussion, the isobars, $^{134}$Sn and $^{134}$Te with A = 134 are pointed out with an asterisk (*).

The structure of N > 82 Te, two protons beyond Z = 50, nuclei are expected to provide a wealth of information on the shell evolution of nuclei at extreme proton-neutron ratios. Particular to these nuclei is the impact of the interactions of the two valence protons with the valence neutrons on the overall shell structure. Below N = 82, Te isotopes exhibit typical vibrational character, where co-existing single-particle and collective structures are manifest. [9-15]. This vibrational character is still present in $^{136}$Te, N = 84 [16], however, the reduction of the energy of its $2^+$ state in comparison to the N < 82 isotopes, as shown in Fig. 1, suggests the onset of a stabilized ground-state deformation [17, 18], which is predicted by Ref. [19, 20] to be prolate. Despite this relatively low $2^+$ energy, a Coulomb excitation study of $^{136}$Te reported an unexpectedly low reduced E2 transition strength [21], which contradicts the predicted deformation. The discrepancy was explained by the quasiparticle random phase approximation as a neutron-pairing reduction [22] and by the Monte Carlo Shell Model as neutron dominance through asymmetric proton-neutron couplings [23]. Furthermore, a large scale shell model calculation also pointed out the importance of the neutron dominance in the wave function of excited states in neutron-rich Te isotopes [24]. A recent study [8] showed that the ratio of the first $4^+$ to $2^+$ energies, $E(4^+)/E(2^+)$ for $^{138}$Te, with N = 86, is identical to that of $^{130}$Te with N = 78. As a result, the energy ratios show a symmetric pattern in Te isotopes with the same valence neutron holes and particles with respect to N = 82. Here we address the following questions; whether, or not the first $2^+$ level energy decreases continuously at N = 88 as it does between N = 84 (607 keV) and at N = 86 (401 keV), and how does the value of $E(4^+)/E(2^+)$ develop at N = 88, i. e. does it remain ~ 2, or does it increase ? To date, however, no experimental data on $^{140}$Te has been published. In this work, we report on the first observation of excited states of $^{140}$Te populated by the β decay of $^{140}$Sb. In addition, we present the β-decay scheme of $^{140}$Sb, including β-delayed one- and two-neutron emission.

The experiment was carried out at the Radioactive Isotope Beam Factory (RIBF) of the RIKEN Nishina Center. The parent nuclides of $^{140}$Te, $^{140}$Sb, were produced by the in-flight fission of a 345 MeV per nucleon $^{238}$U beam on a $^9$Be target and selected by the first stage of the BigRIPS separator [25]. The mean intensity of the primary beam was 5 to 7 pnA over the course of the five days of beam time. Fission fragments, transported through the zero-degree spectrometer (ZDS), were unambiguously identified by the Bρ-ΔE-time-of-flight method [26]. They were implanted into the wide-range active silicon strip stopper array for beta and ion detection (WAS3ABi), which comprised five layers of 1-mm-thick double-sided silicon-strip detectors (DSSSDs) [27]. It was surrounded by two 2 mm thick plastic scintillators for the rejection of charge-exchange reactions. During the beam time, a total of $7.8\times10^3$ ions for $^{140}$Sb were collected among about $10^7$ total ions. Emitted γ rays, following the β decay of $^{140}$Sb were then detected by the EUROBALL-RIKEN high-purity germanium (HPGe) Cluster array (EURICA) [28] surrounding WAS3ABi. See Refs. [8, 28, 29] for more details of experimental methods. Figure 2 shows the β-delayed γ-ray spectrum of $^{140}$Sb, where the ion-β time was limited to 450 ms. The broad peaks around 425 keV in this singles spectrum are visible. They are composed of triple photo-peaks as shown in the inset of Fig. 2; 423, 425, and 428 keV. The singles spectra obtained by various elapsed time selection allowed us determination for the γ-ray transitions associated with the daughter nuclei; $^{140}$Te, $^{139}$Te (β-delayed one-neutron emission), and $^{138}$Te (β-delayed two-neutron emission).

The results from the γ-γ coincidence analysis are shown in Fig. 3, where the 423- and 425-keV transitions are shown to be in mutual coincidence, no other transitions were correlated with the 423- and 425-keV transitions. In contrast, the 428 keV peak turned out to be independent of the 423- and 425-keV transitions. On the basis of the γ-γ coincidence data and γ-ray intensities in the singles spectra, we propose that the 423- and 425-keV peaks should be assigned as γ-ray transitions in $^{140}$Te. Based on the fitting results for various spectra under different ion-β time conditions, as an example shown in the inset of Fig. 3(a), we adopted the intensity of the 423 keV and the 425 keV to be 100(16) and 89(16) %, respectively. Accordingly, the 425- and the 423-keV transitions are assigned as the $4^+$ to $2^+$ and the $2^+$ to $0^+$ levels, respectively in $^{140}$Te. We show also in the inset of Fig. 3(b) the timing spectrum associated with the 423-425-keV γ-ray peaks on the basis of γ-time matrix. The quoted decay half-life was determined using a single-component exponential decay with a least-square fit minimization method and assuming a constant background level. As indicated in Fig. 3, the decay half-life was measured to be 124(30) ms when gating on the 423- and 425- keV transitions. Additionally, the 428-keV peak showed a similar time-decay curve but too low in yields to extract a half-life. We emphasize that in the present data there are no delayed γ-rays indicating isomers with a few ns or above in $^{140}$Te.

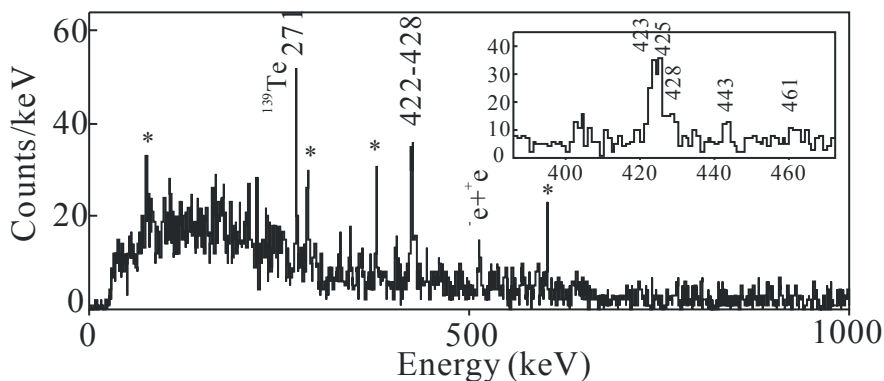

Fig. 2 Singles gamma-ray spectrum associated with the β decay of $^{140}$Sb obtained in 450 ms time interval after ions are implanted on the active target. The inset is a zoom spectrum in representing the 380 to 470 keV region. Peaks with an asterisk are room- and beam-induced backgrounds from random coincidence with β events.

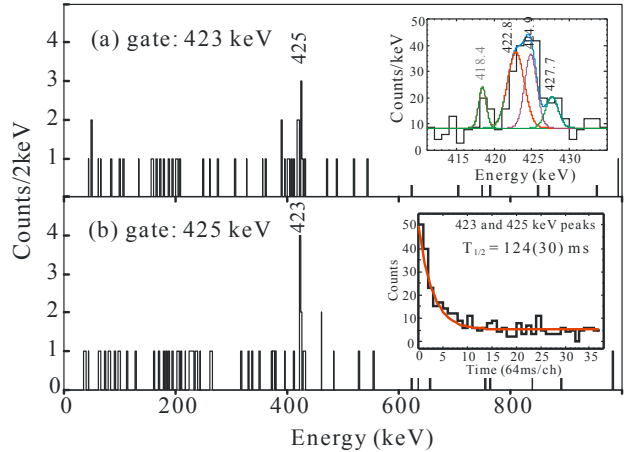

Fig. 3. Coincidence γ-rays gated on (a) 423 keV and (b) 425 keV, which are assigned to the transitions in $^{140}$Te. The inset of (a) shows their individual peaks including a Gaussian fit. The inset of (b) shows a timing spectrum gated on the 423- and 425-keV photo-peaks. The solid line represents the fitting using an exponential decay curve. The number in parenthesis is the error in the last digit.

The 428-keV transition should be assigned to $^{139}$Te rather than $^{140}$Te, since the 428-keV excitation energy, close to the value of the 2$^+$ state of $^{140}$Te, excludes a possibility that this peak is a transition to the ground state in $^{140}$Te. Instead, this γ-ray is most likely to be a transition from a state with $J^\pi = 11/2^-$ to the ground state with $J^\pi = 7/2^-$ in $^{139}$Te. We notice that the transition energies of 1180 (608) keV in $^{135}$Te ($^{137}$Te) from the 11/2$^-$ state to the 7/2$^-$ ground state are close to the first 2$^+$ excited energies of 1279 (607) keV in $^{134}$Te ($^{136}$Te). This feature indicates that transitions from the 11/2$^-$ state to the 7/2$^-$ ground state in odd Te are due to core excitations in the neighboring even-even isotopes. In turn, the 428-keV transition is also found to be similar in energy to the 461-keV transition between the 2$^+$ state and the ground state in $^{138}$Te. Following this systematic trend, we can draw a conclusion that the 428 keV corresponds to the core transition coupled to a neutron depopulating the 11/2$^-$ state to the 7/2$^-$ ground state in $^{139}$Te. The strong 271-keV peak observed in our data is known to be a transition from the 9/2$^-$ state to the ground state in $^{139}$Te. It is important to know that the 428-keV transition could not be populated (or too weak to be observed), though the 271-keV transition was strongly populated, in the decay from the 7/2$^+$ ground state of $^{139}$Sb. The intensity of both γ rays, 271 keV and 428 keV, in $^{139}$Te was found to be about 59 % with respect to the total γ-ray intensity from the decay of $^{140}$Sb. In addition, we found weak γ-rays belonging to transitions in $^{138}$Te, 461 keV and 443 keV [8], as shown in the inset of Fig. 2. We assigned these two transitions to $^{138}$Te produced by β-delayed two-neutron (2n) emission. This β-2n branching ratio, based on γ-ray intensities, was found to be about 5 %. We emphasize that there is no strong evidence of direct feeding from the β-decay of $^{140}$Sb to the ground state of $^{140}$Te from a comparison of the number of implants and the associated with β–γ ray coincident events. However, in cases of β-delayed one and two-neutron emission, such a branch cannot be ruled out. Therefore the probability of the β-delayed neutron-emission may be more intense than that obtained in the present work. According to the feeding pattern observed in the present work, we can restrict the possible spin-parity values of the ground state of $^{140}$Sb, since β-decay populates excited states up to 4$^+$ in $^{140}$Te, implying that the ground-state spin is most likely either 3 and 4. In addition, the observation of γ-ray transitions in $^{139}$Te produced by β-delayed neutron emission provides stringent constraints on the spin-parity assignment. Taking into consideration the 271-keV transition depopulating the 9/2$^-$ state and the 428-keV transition depopulating the (11/2$^-$) state in $^{139}$Te, the most likely spin-parity of $^{140}$Sb is 4$^-$, rather than 3$^-$. A comparable intensity of the 425- and 423-keV transitions supports this result. A possible β-decay of $^{140}$Sb to $^{140}$Te is expected to stem from, primarily, the conversion of a neutron in the $f_{7/2}$ orbital into a proton in the $g_{7/2}$ (or $d_{5/2}$) orbital; such conversions are first-forbidden Gamow-Teller transition. Therefore, the negative parity assignment is based on a proton-neutron configuration of $\pi g_{7/2}$ (or $d_{5/2}$) $\nu f_{7/2}$. Furthermore, the deduced log(ft) value, 6.3(2) is in reasonable agreement with our assignment $4^-$. The present results for the decay scheme of $^{140}$Sb are summarized in Table I and the resultant decay scheme of $^{140}$Sb is shown in Fig. 4.

Table I. Summary of the β decay of $^{140}$Sb to $^{140}$Te, $^{139}$Te by one-neutron emission, and $^{138}$Te by two-neutron emission. Probabilities for the respective decay branch are based on γ-ray measurements. Energies are given in keV. The errors include statistical error of each transition and systematical error 10 % of the detection efficiency. The numbers in parentheses are the errors in the last digit.

|  | log(ft)* | β-delayed probability* | Observed γ rays ; relative intensities | Spin-parity assignment for γ-ray transitions |
|---|---|---|---|---|
| $^{140}$Sb |  |  |  | $T_{1/2}$ = 124(30) ms, $J^\pi$ = $(4^-)$ for ground state |
| $^{140}$Te | 6.3(2) | 5.4(9) % | 422.8(7); 24(6) <br> 424.9(6); 23(6) | $2^+ \to 0^+$ <br> $4^+ \to 2^+$ |
| $^{139}$Te |  | 6.1(10) % <br> 3.0(10) % | 271.3(5); 30(3) <br> 427.7(7); 12(2) | $9/2^- \to 7/2^-$ <br> $(11/2^-) \to 7/2^-$ |
| $^{138}$Te |  | 1.3(7) % | 460.8(5); 5(1) <br> 442.8(5); 6(1) | $2^+ \to 0^+$ <br> $4^+ \to 2^+$ |

* considering the number of $^{140}$Sb implanted on the active target to be 7,833 and the Q value is 12420 keV.

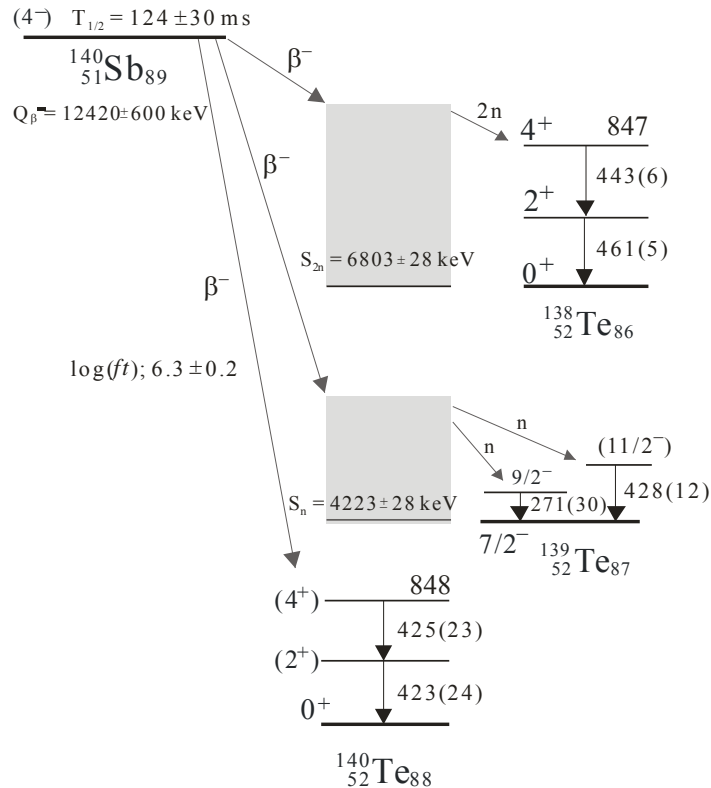

Fig. 4. The β decay scheme of $^{140}$Sb deduced from the present work. The intensities of the γ-ray transitions in each daughter Te nuclei are normalized to per 100 $^{140}$Sb decay. The $Q_\beta$, $S_n$ and $S_{2n}$ values are quoted from [30]. The numbers in parentheses are the relative γ-ray intensities.

The starting point for the discussion is the description of the systematic behavior of low-lying level properties in the even-even nuclei of interest. A highly illuminating observable which adheres to systematic behavior is the excitation energy ratio of the first $2^+$ and $4^+$ states, $R = E(4^+)/E(2^+)$. This value evolves from < 2 for a spherical nucleus through 2 for a vibrator, to 3.33 for a deformed axial rotor [31, 32]. Figure 5(a) illustrates the systematics of R values for a given neutron number (isotones) along Z = 50 (Sn) to Z = 70 (Yb), while Fig. 5(b) demonstrates their differences, ΔR, between a pair of isotopes with the same number of valence neutron holes and particles with respect to the N = 82 closed shell. Accordingly, ΔR (88-76) means the R values difference between, as an example, $^{140}$Te with N = 88 and $^{128}$Te with N = 76.

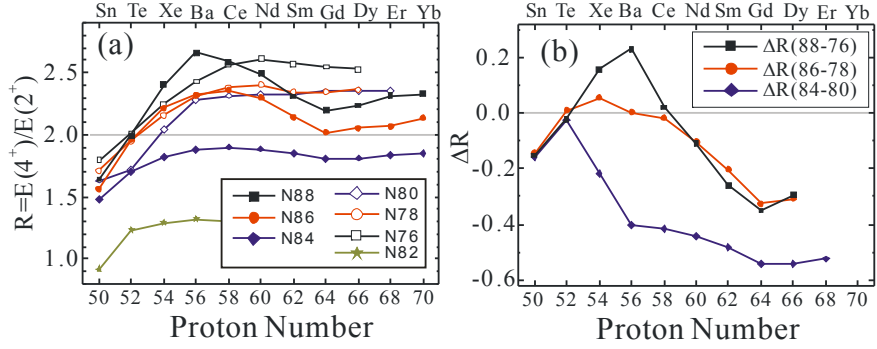

Fig. 5. Systematic plots for (a) R (= $E(4^+)/E(2^+)$) values as a function of proton number in isotopes between N = 76 and 88 and (b) ΔR correlations for a pair of isotopes with the same number of valence neutron holes and particles with respect to the closed shell N = 82.

From the R values, we notice the following characteristics: First, for isotones above N = 82 a *pseudo subshell* within 50 ≤ Z ≤ 64 is formed, it is strongly reinforced with N = 88, moderately with N = 86, and weakly with N = 84. It should be noted that this pseudo subshell has a total spin state j = 13/2 with capacitance of 14 nucleons occupancy. In contrast, there is no such a subshell responding to neutron numbers below N = 82. Secondly, the R values of Te isotopes are centered about two values 2.0 (N = 76, 78, 86, 88) and 1.75 (N = 80, 84). This is quite a striking result that is not expected by the present theoretical, or empirical predictions, for instance, see Ref. [24] in which $^{140}$Te was described like a triaxial rotator with R = 2.33. The characteristics in the ΔR systematics are as follows: for Te there is little difference in the R values, giving ΔR ~ 0; for Sn, the differences are constant at ΔR ~ - 0.175; for Xe and Ba, the values split into three regimes; positive, close to zero, and negative; for Ce ( Z = 58) to Dy ( Z = 66) the lines have two branches which extend into the negative region, and, finally, for N = 88 there is a distinctive peak at Z = 56 and a pronounced valley at Z = 64. This feature provides further confirmation of the existence of a pseudo-subshell between Z = 50 and Z = 64, strongly reinforced by N = 88. The above phenomenological arguments indicate for Te there is no evidence of the associated nuclear structure change with deformation between N = 76 and N = 88. In other words, Te isotopes give a symmetrical signature that the same valence space results in a similar collectivity: *the valence neutron symmetry*. Moreover, an emergence of only negative values of ΔR indicates that Sn isotopes above N = 82 are less deformed than those below N = 82. The Ba isotopes show a greater deformation with N = 88, while $^{140}$Ba, with N = 84, is less deformed than their counter-part below N = 82, and $^{142}$Ba with N = 86 has the same in deformation as $^{134}$Ba with N = 78. The Xe isotopes follow similar systematics. It is worthwhile to emphasize that the pseudo subshell has a capacity of 14 protons (7 pairs of protons). Given that half-filled, high-j orbitals drive nuclear deformation, in the presence of a 50 ≤ Z ≤ 64 pseudo-subshell, Ba (Z = 56) and Ce (Z = 58) are expected to have maximal deformations. As shown in Fig. 6, this assumption explains why $^{144}$Ba and $^{144}$Ce have a maximum value of R at N = 88 and at N = 86, respectively.

The reduction of neutron pairing can be explained in terms of the large difference between the proton and neutron gap in $^{134}$Te$_{82}$ and $^{134}$Sn$_{84}$, see Fig. 1 where they are denoted by a star symbol. Such features for the neutron-rich nuclei above N = 82 are also illuminated by the energy difference between the 2$^+$ states in a pair of isotopes above and below N = 82. Figure 6 describes the systematics based on such an energy difference, ΔE(2$^+$), between those with the same neutron holes below N = 82 and neutron particles above N = 82. We found, for example, the ΔE(2$^+$) values for a pair of isotopes for Sn versus Te nuclei between N = 80 and 84, N = 78 and 86, and N = 76 and 88 are 495 : 368, 481 : 379, and 426 : 320 in keV, respectively: demonstrating *the valence neutron asymmetry*. As a result, the different pairing properties for Sn, Te, Xe, and Ba appear to be dependent on the ratios of the valence protons and the valence neutrons with respect to N = 82. Hence, the first 2$^+$ states in neutron-rich nuclei above N = 82 are expected to have proton-neutron mixed asymmetric interactions as a pair of neutrons distributes dominantly over a pair of protons.

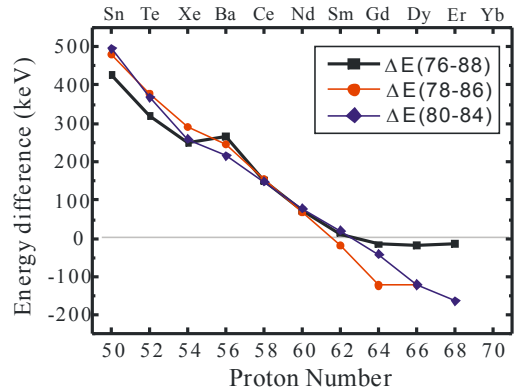

Fig. 6. Systematic plots for the 2$^+$ states energy differences as a function of proton number between a pair of isotopes with the same number of valence neutron particles and holes with respect to the closed shell N = 82.

Neutron-dominance in 2$^+$ state leads to a weaker B(E2) value as already pointed out in $^{136}$Te [21]. Interestingly, the ΔE(2$^+$) values are almost identical within a variation of 10 % for both Sn and Te. This result implies that the neutron pairings at N = 86 and 88 are likely comparable to that at N = 84. Contrariwise, from the ΔR values, we notice that Te isotopes maintain their collective character, showing a typical vibrator with ΔR ≈ 0 over the N = 82 region. We raise here the following questions; what number of neutrons relative to a proton pair contribute, in energy, to the first 2$^+$ state, and how does the neutron dominance in the 2$^+$ state effect the B(E2) strength. A sophisticated shell model theory is required for an understanding of the underlying physics in $^{140}$Te, employing the proton-neutron mixed asymmetric interactions.

In conclusion, we provide the first data of the β-decay scheme of the very neutron-rich $^{140}$Sb and the excited states in of its daughter nucleus, $^{140}$Te. The half-life and spin-parity of $^{140}$Sb were measured to be 124(30) ms and (4$^-$), respectively. We identified β-delayed, β-delayed one-neutron emission, and β-delayed two-neutron emission from the decay of $^{140}$Sb and determined their decay probabilities on the basis of γ-ray measurements involved in each daughter nucleus. We deduced successfully the first 2$^+$ and 4$^+$ excited states in $^{140}$Te and assigned a new transition in $^{139}$Te. The present data provided a clear evidence that the character of $^{140}$Te persists as a typical vibrator, having E(4$^+$)/E(2$^+$) ~ 2 seen in other isotopes: demonstrating *the valence neutron symmetry*. We discussed the level structure of $^{140}$Te based on the systematics of the 2$^+$ and 4$^+$ states in the vicinity of N = 82. Along with the study of Te isotopes, we addressed some interesting aspects by focusing on the distinctive features of this region, namely *the valence neutron symmetry and asymmetry* in isotopes from Sn (Z = 50) to Yb (Z = 70).

This work was carried out at the RIBF operated by RIKEN Nishina Center, RIKEN and CNS, University of Tokyo. We acknowledge the EUROBALL Owners Committee for the loan of germanium detectors and the PreSpec Collaboration for the readout electronics of the cluster detectors. Part of the WAS3ABi was supported by the Rare Isotope Science Project, which is funded by the Ministry of Science, ICT, and Future Planning (MSIP) and the National Research Foundation (NRF) of Korea. This research was partly supported by JSPS KAKENHI Grant No. 25247045 and the National Research Foundation Grant funded by the Korean Government (Grants No.NRF-2009-0093817 and No.NRF-2013R1A1A2063017). FR-JP LIA support is also acknowledged.